\documentclass[11pt]{revtex4}
 
\usepackage{relsize} 
\usepackage{hyperref}
\usepackage{amsmath,amsfonts,amssymb}
\hypersetup{dvips,dvipdfm,colorlinks=true,urlcolor=magenta,filecolor=magenta,linktoc=page,citecolor=red,linkcolor=blue,bookmarks=true}
\usepackage{graphicx,epstopdf}
 \usepackage[inline]{enumitem}
 
\begin{document}

\title{Does emergent scenario in Ho\v{r}ava-Lifshitz Gravity demand a ghost field?}

\author{Akash Bose$^1$\footnote {bose.akash13@gmail.com}}
\author{Subenoy Chakraborty$^1$\footnote {schakraborty.math@gmail.com}}
\affiliation{$^1$Department of Mathematics, Jadavpur University, Kolkata-700032, West Bengal, India}
\begin{abstract}
	The non-singular model of the Universe i.e. emergent scenario is now very well known in cosmology. In Einstein gravity such type of singularity free solution is possible in the context of non-equilibrium thermodynamical prescription (both in first and 2nd order theory) with particle creation mechanism. Also there are various models of emergent scenario in different gravity theories. The present work is a twofold attempt namely (i) it has been examined whether it is possible to have singularity free solution in Ho\v{r}ava-Lifshitz modified gravity theory, (ii) if such singularity free solution exists then whether such solution is thermodynamically viable. It is found that perfect fluid matter distribution with phantom nature can generate such type of solution for closed and flat model. Finally thermodynamical analysis has been studied both at the apparent horizon and at any arbitrary horizon.
\end{abstract}
\maketitle
$~~~~~~$Keywords: Ho\v{r}ava-Lifshitz gravity; emergent scenario; thermodynamical equilibrium.
\section{Introduction}
The Big Bang singularity is a consequence of the fact that Einstein's gravity is a classical field theory. The proposal of cosmologists to overcome this singularity can be classified as bouncing universe or the emergent universe which is very much relevant in the context of inflationary models. An emergent Universe \cite{Harrison:1967zz}-\cite{Chakraborty:2014ora} is a model of the Universe without any time-like singularity, existing throughout the time axis (i.e. $-\infty<t<+\infty$) and static in nature in the infinite past (t$\rightarrow-\infty$).

In this non-singular model, the initial Big Bang singularity is replaced by an Einstein static era (with $a=a_0\neq0$) so that physical quantities namely energy density, pressure etc$.$ all are finite at the beginning. Subsequently, the Universe smoothly expands (known as pre-inflationary era) and enters into inflationary epoch. Then as usual in standard cosmology there is a phase of reheating and gradually the Universe approaches to the usual classical thermal radiation dominated era of evolution. The following table \ref{t1} shows the behaviour of the scale factor and Hubble parameter in the above eras of evolution: \cite{delCampo:2011aa}
\begin{table}[h]
	\caption{\label{t1}Nature of the scale factor and Hubble parameter at different phases of evolution}
	\begin{tabular}{|c|c|c|}
	\hline
	Name of the epoch&Scale factor&Hubble parameter\\\hline
	Emergent era&$a=a_0(\neq 0)$&$H=0$\\\hline
	Pre-inflationary era&`$a$' increases from $a_0$ to $a_i$ $(>a_0)$&$H$ increases from zero to a finite value $H_0$\\\hline
	Inflationary era&`$a$' increases exponentially with $a_i<a<a_f$&$H=H_0$\\\hline
	Radiation era&$a>a_f$&$H$ decreases gradually from $H_0$\\\hline
	\end{tabular}
\end{table}
	
From observational point of view, cosmologists are giving much attention to the modified theories of gravity over recent years due to geometric matter component, which may be considered among the most natural choices for dark energy. In the present work, such a modified gravity theory namely Ho\v{r}ava-Lifshitz (HL) gravity is considered for constructing an emergent model of the Universe. The HL gravity theory is a power-counting renormalization theory with consistent ultra-violet behavior. Also it provides Einstein gravity as a critical point. Using the detailed-balance condition the gravitational action integral for HL gravity can be written as \cite{Kiritsis:2009sh}-\cite{Horava:2009uw}
\begin{eqnarray}
A_{HL}&=&\int dtd^3x\sqrt{g}N\bigg[\frac{2}{\kappa^2}(K_{ij}K^{ij}-\lambda K^2)+\frac{\kappa^2}{2w^4}C_{ij}C^{ij}-\frac{\kappa^2\mu}{2w^2}\frac{\epsilon^{ijk}}{\sqrt{g}}R_{il}\nabla_j R^l_k+\frac{\kappa^2\mu^2}{8}R_{ij}R^{ij}\nonumber\\&~&-\frac{\kappa^2\mu^2}{8(3\lambda-1)} \left(\frac{1-4\lambda}{4}R^2 + \Lambda R- 3\Lambda^2\right)\bigg]\label{eq1}
\end{eqnarray}

where $N$ is the Lapse function, $K_{ij}$ is the extrinsic curvature tensor, $\epsilon^{ijk}$ is the totally antisymmetric tensor, $\lambda$ is a dimensionless constant, the quantities $w$ and $\mu$ are purely constants and $C^{ij}$, the Cotton tensor is given by \cite{Horava:2009uw}
\begin{equation}
C^{ij}=\frac{\epsilon^{ikl}}{\sqrt{g}}\nabla_k\left(R^j_l-\frac{1}{4}R\delta^j_l\right)\label{eq2}
\end{equation}



There is a nice overview of HL gravity theory in \cite{Sotiriou:2010wn}, distinguishing between various versions of it. Also the dynamics and the viability of these versions has been discussed in this review. Further in \cite{Luongo:2018oil} the authors have investigated the limits imposed by observations on the minimal paradigm of HL in the background of homogeneous and isotropic space-time. They have found statistical inconsistencies of the model, ruling out HL paradigm at the level of background cosmology. The validity of HL cosmology both at early and late times is yet to be analyzed and they have speculated that by adding new extra terms into the Lagrangian one may have viable HL gravity model.
\section{Emergent scenario in Ho\v{r}ava-Lifshitz gravity}
The modified Einstein field equations for perfect fluid in the frame work of Ho\v{r}ava-Lifshitz gravity theory obtained by varying the above action (\ref{eq1}) with respect to the metric are the following non-linear differential equations (choosing $\kappa^2=8\pi G=1,~N=1$)\cite{Leon:2019mbo}
\begin{eqnarray}
3H^2&=&\frac{1}{2\left(3\lambda-1\right)}\rho-\frac{3\mu^2}{16(3\lambda-1)^2}\left(\Lambda-\frac{k}{a^2}\right)^2\label{eq4}\\
2\dot{H}+3H^2&=&-\frac{1}{2\left(3\lambda-1\right)}p-\frac{\mu^2}{16(3\lambda-1)^2}\left(\Lambda-\frac{k}{a^2}\right)\left(3\Lambda+\frac{k}{a^2}\right)\label{eq5}
\end{eqnarray}

where $k$, the curvature scalar takes the values $-1$, $0$, $+1$ according as open, flat and closed model while the continuity equation for the perfect fluid is
\begin{equation}
\dot{\rho}+3H(p+\rho)=0\label{eq6}
\end{equation}
with barotropic equation of state : $p=\omega\rho$.


Now eliminating $\rho$ between the field equations (\ref{eq4}) and (\ref{eq5}) (using the above equation of state) one gets the cosmic evolution equation as
\begin{equation}
2\dot{H}=-3(1+\omega)H^2-\frac{\mu^2}{16(3\lambda-1)^2}\left(\Lambda-\frac{k}{a^2}\right)\left[3\Lambda(1+\omega)+\frac{k}{a^2}(1-3\omega)\right]\label{eq7}
\end{equation}

In order to examine the possibility of a singularity free cosmological solution (i.e. emergent scenario) in the present modified gravity theory one has to take into account of the following asymptotic behavior for the Hubble parameter and scale factor as: \cite{Chakraborty:2014ora}
\begin{eqnarray}\label{eq8}
(i)&&a\rightarrow a_0~,~H\rightarrow 0~\mbox{as}~t\rightarrow-\infty~;\nonumber\\
(ii)&&a\simeq a_0~,~H\simeq 0~\mbox{for}~t\ll t_0~;\nonumber\\
(iii)&&a\simeq a_0 \exp\{H_0(t-t_0)\}~,~H\simeq H_0~\mbox{for}~t\gg t_0~.
\end{eqnarray}
where $a_0$, a constant is the tiny value of the scale factor during emergent era of evolution, and similarly $H_0(>0)$ is the value of the Hubble parameter at $t=t_0$.

From the above asymptotic behavior of the scale factor one can have the explicit form of the scale factor as \cite{Debnath:2008nu},\cite{Chakraborty:2014ora}
\begin{equation}\label{eq9}
a=a_0\left[1+e^{\frac{H_0}{\alpha}(t-t_0)}\right]^\alpha
\end{equation} 

so that\vspace{-.5cm}
\begin{equation}
H=H_0\left[1-\left(\frac{a_0}{a}\right)^\frac{1}{\alpha}\right],\label{eq10}
\end{equation}

and\vspace{-.5cm}\begin{equation}
\alpha a\frac{dH}{da}=H_0\left(1-\frac{H}{H_0}\right)\label{eq11}
\end{equation}

Now using the evolution equation (\ref{eq11}) for the Hubble parameter into equation (\ref{eq7}) one has
\begin{eqnarray}\label{eq12}
-(1+\omega)=\frac{2aH\dfrac{dH}{da}+\dfrac{\mu^2k}{4a^2(3\lambda-1)^2}\left(\Lambda-\dfrac{k}{a^2}\right)}{3H^2+\dfrac{3\mu^2}{16(3\lambda-1)^2}\left(\Lambda-\dfrac{k}{a^2}\right)^2}
\end{eqnarray}

and the energy density of the perfect fluid has the expression
\begin{equation}\label{eq13}
\rho=6H^2(3\lambda-1)+\frac{3\mu^2}{8(3\lambda-1)^2}\left(\Lambda-\frac{k}{a^2}\right)^2
\end{equation}

%
From the above expressions for equation of state parameter and energy density, one may conclude the following:

\begin{enumerate*}[label=(\alph*)]
\item For flat model of the Universe (i.e. $k=0$) to have a singularity free era of evolution at the early phase for the Ho\v{r}ava-Lifshitz modified gravity theory, the perfect fluid is at the phantom barrier $(\omega=-1)$ i.e. behaves as cosmological constant. However, phantom fluid $(\omega<-1)$ is needed for the pre-inflationary era.\\

\item For closed model (i.e. $k=+1$) of the Universe, if $a_0>\dfrac{1}{\Lambda}$ and also $a_0>l_{pl}$ (Planck length) (to avoid quantum gravity effect) i.e. $a_0>\min\left(\dfrac{1}{\Lambda},l_{pl}\right)$ then phantom perfect fluid is needed both for emergent era and in pre-inflationary epoch. Here the cosmological constant has no role in characterizing the nature of the perfect fluid.\\

\item In an open model (i.e. $k=-1$) of the Universe, normal perfect fluid describes both emergent as well as pre-inflationary phases of evolution.
\end{enumerate*}

Further using (\ref{eq10}) and (\ref{eq11}) one has the energy density and equation of state parameter as explicit form in terms of the scale factor `$a$' as
\begin{eqnarray}
\rho(a)&=&6H_0^2(3\lambda-1)\left[1-\left(\frac{a_0}{a}\right)^\frac{1}{\alpha}\right]^2+\frac{3\mu^2}{8(3\lambda-1)^2}\left(\Lambda-\frac{k}{a^2}\right)^2\label{eq14}\\
-3(1+\omega)&=&\frac{\dfrac{2H_0^2}{\alpha}\left[1-\left(\dfrac{a_0}{a}\right)^\frac{1}{\alpha}\right]\left(\dfrac{a_0}{a}\right)^\frac{1}{\alpha}+\dfrac{\mu^2k}{4a^2(3\lambda-1)^2}\left(\Lambda-\dfrac{k}{a^2}\right)}{H_0^2(3\lambda-1)\left[1-\left(\dfrac{a_0}{a}\right)^\frac{1}{\alpha}\right]^2+\dfrac{\mu^2}{16(3\lambda-1)^2}\left(\Lambda-\dfrac{k}{a^2}\right)^2}\label{eq15}
\end{eqnarray}

\begin{figure}[h]
	\begin{minipage}{.495\textwidth}
		\centering\includegraphics[width=.95\textwidth]{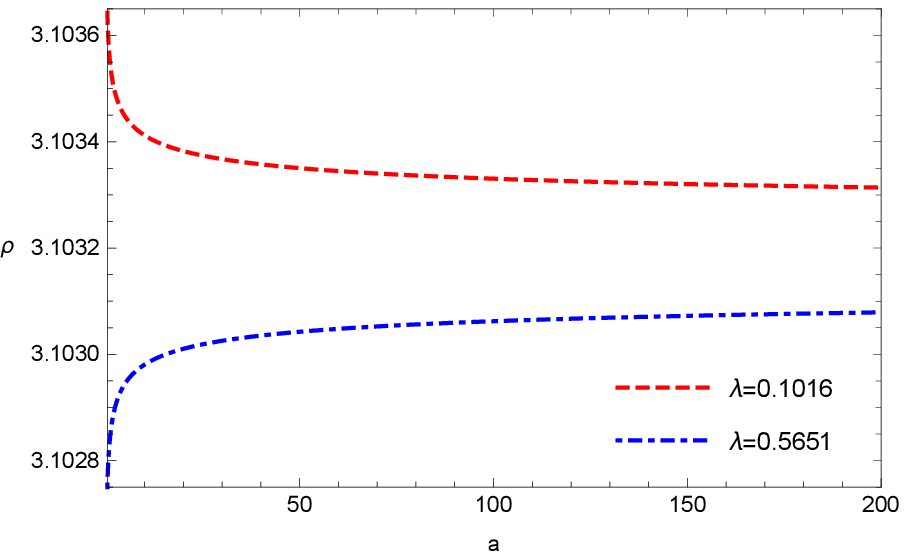}
		{$k=0,\Lambda\neq 0$}
	\end{minipage}
	\begin{minipage}{.495\textwidth}
	\centering\includegraphics[width=.95\textwidth]{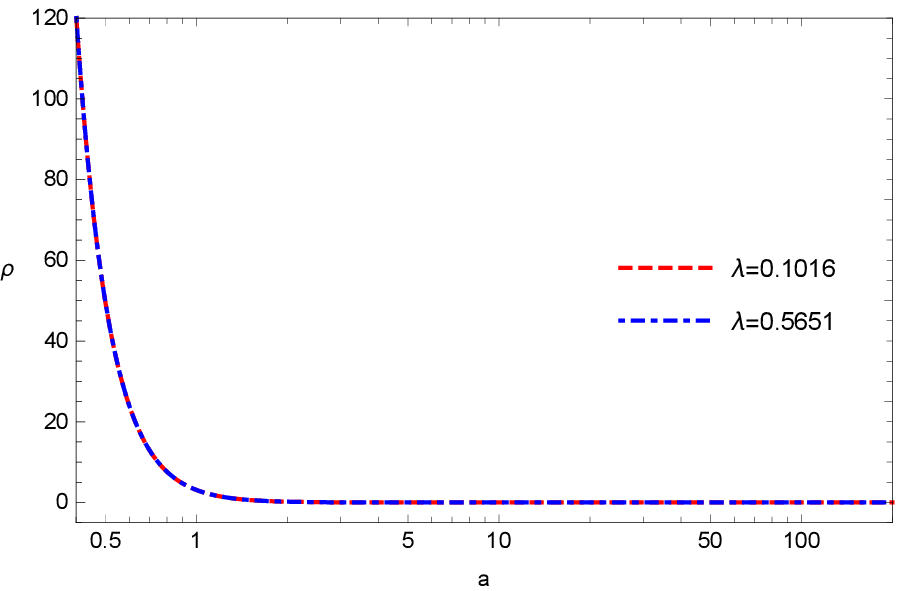}
	{$k\neq0,\Lambda= 0$}
\end{minipage}
	\begin{minipage}{.495\textwidth}
		\centering\includegraphics[width=.95\textwidth]{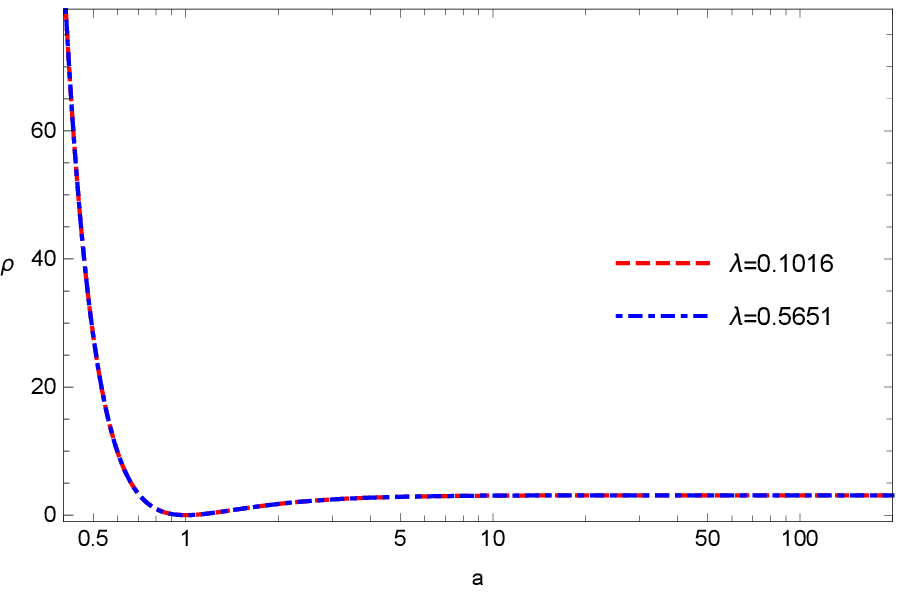}
		{$k=1,\Lambda\neq 0$}
	\end{minipage}
	\begin{minipage}{.495\textwidth}
		\centering\includegraphics[width=.95\textwidth]{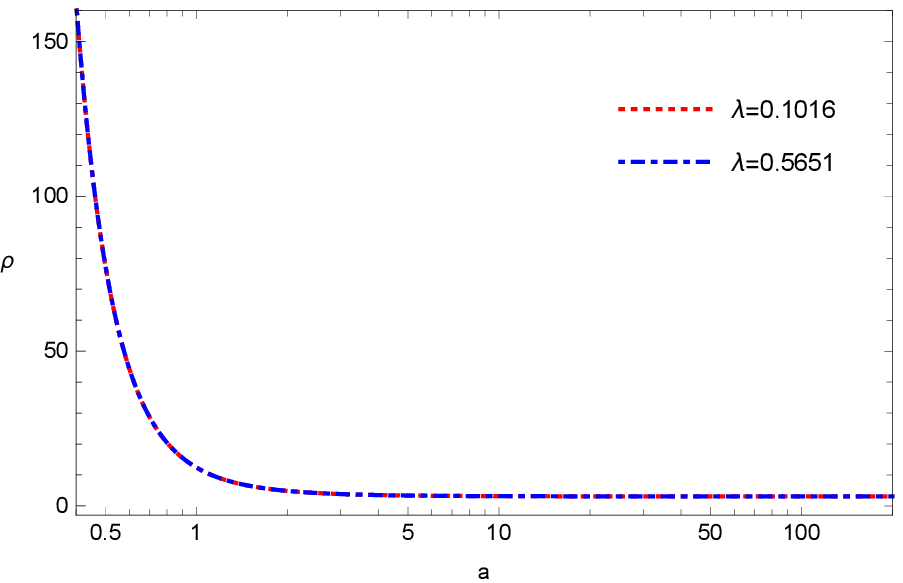}\\
		{$k=-1,\Lambda\neq0$}
	\end{minipage}
\caption{Energy density $\rho$ is plotted against scale factor with $a_0$=0.4, $H_0$=.01, $\alpha$=4,  $\mu$=2}
\label{fig1}
\end{figure}

\begin{figure}[h]
\begin{minipage}{.495\textwidth}
	\centering\includegraphics[width=1\textwidth]{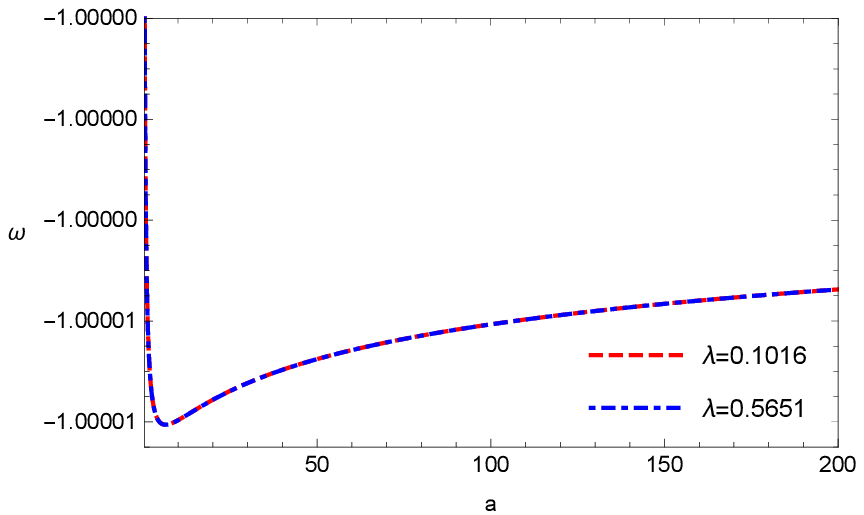}
	{$k=0,\Lambda\neq 0$}
\end{minipage}
\begin{minipage}{.495\textwidth}
	\centering\includegraphics[width=.95\textwidth]{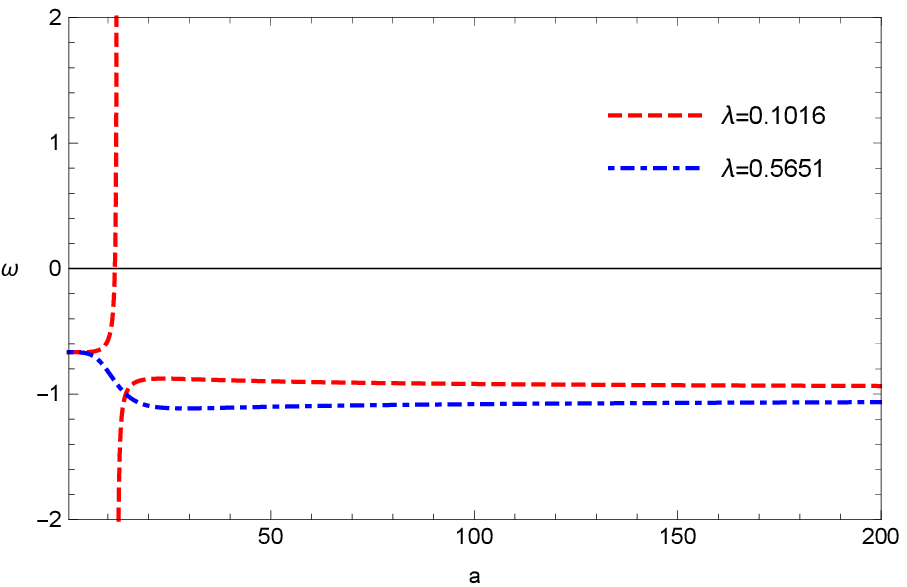}
	{$k\neq0,\Lambda= 0$}
\end{minipage}
\begin{minipage}{.495\textwidth}
	\centering\includegraphics[width=.95\textwidth]{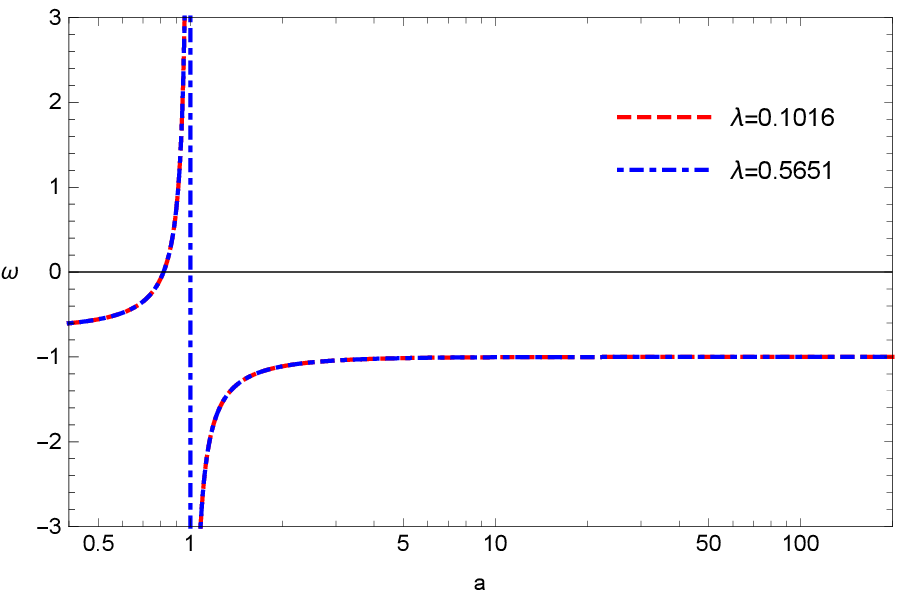}
	{$k=1,\Lambda\neq 0$}
\end{minipage}
\begin{minipage}{.495\textwidth}
	\centering\includegraphics[width=.95\textwidth]{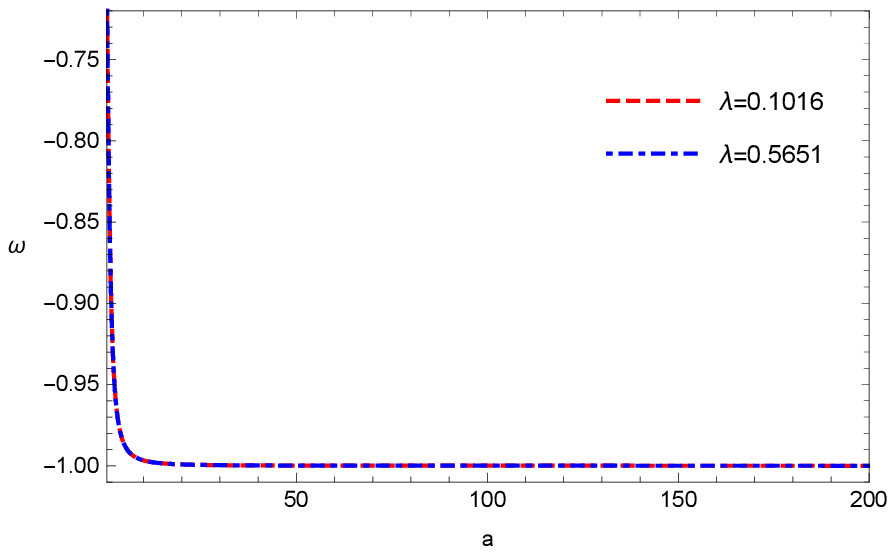}\\
	{$k=-1,\Lambda\neq0$}
\end{minipage}
\caption{Equation of state parameter $\omega$ is plotted against scale factor with $a_0$=0.4, $H_0$=.01, $\alpha$=4,  $\mu$=2}
\label{fig2}
\end{figure}

In figure \ref{fig1} and \ref{fig2}, the variation of $\rho$ (in equation (\ref{eq14})) and the equation of state parameter (in equation (\ref{eq15})) with the scale factor has been shown for the choices
	 \begin{enumerate*}[label=(\roman*)]
		\item$k=0,~\Lambda\neq0$,
		\item$k\neq0,~\Lambda=0$,
		\item$k\neq0,~\Lambda\neq0$
	\end{enumerate*} respectively, for various choices of the parameters involved.
 
From field theoretic point of view, the perfect fluid has an analogous scalar field description. The action integral for the present gravity mode of a self interacting scalar field takes the form
\begin{equation}
A_m=\int dt  \left[\frac{3\lambda-1}{4}\dot{\phi}^2-V(\phi)\right]a^3\label{eq3}
\end{equation}

The energy density and pressure of the perfect fluid is related to the scalar field as
\begin{eqnarray}
\rho&=&\frac{3\lambda-1}{4}\dot{\phi}^2+V(\phi)\nonumber\\
\mbox{and~~~~}p&=&\frac{3\lambda-1}{4}\dot{\phi}^2-V(\phi)
\end{eqnarray}

Further, for the flat case (i.e. $k=0$) a typical analytic form of the scalar field and the potential as a function of scalar field has the following expression:\begin{eqnarray}
\phi&=&-4i\sqrt{2\alpha}\sin^{-1}\left(\left(\frac{a_0}{a}\right)^\frac{1}{2\alpha}\right)\label{eq16}\\
V(\phi)&=&2(3\lambda-1)H_0^2\bigg[3\cosh^4\left(\frac{\phi}{4\sqrt{2\alpha}}\right)-\frac{1}{\alpha}\cosh^2\left(\frac{\phi}{4\sqrt{2\alpha}}\right)\sinh^2\left(\frac{\phi}{4\sqrt{2\alpha}}\right)+\frac{3\mu^2\Lambda^2}{16(3\lambda-1)^2H_0^2}\bigg]\label{eq17}
\end{eqnarray}
\begin{figure}[b]
	\begin{minipage}{.495\textwidth}
		\centering\includegraphics[width=.85\textwidth]{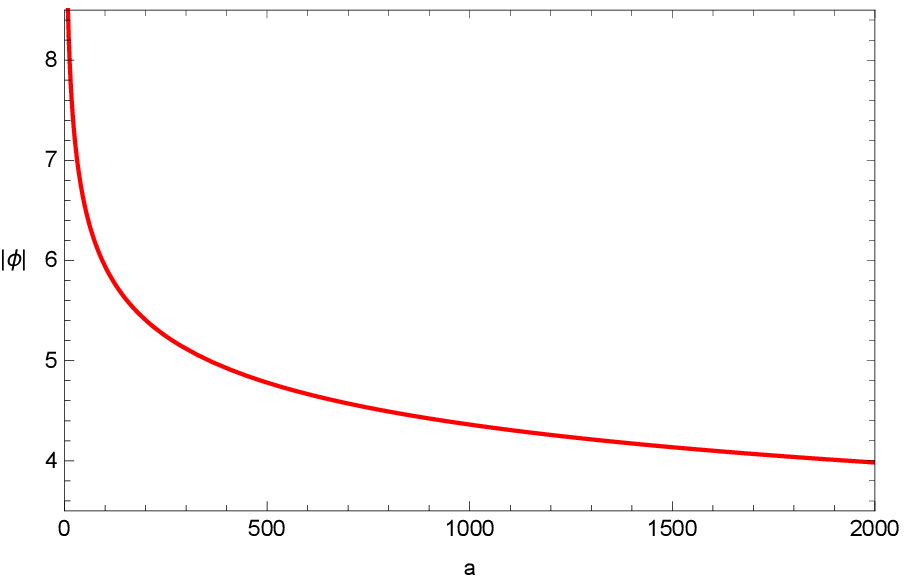}
	\end{minipage}
	\begin{minipage}{.495\textwidth}
		\centering\includegraphics[width=.85\textwidth]{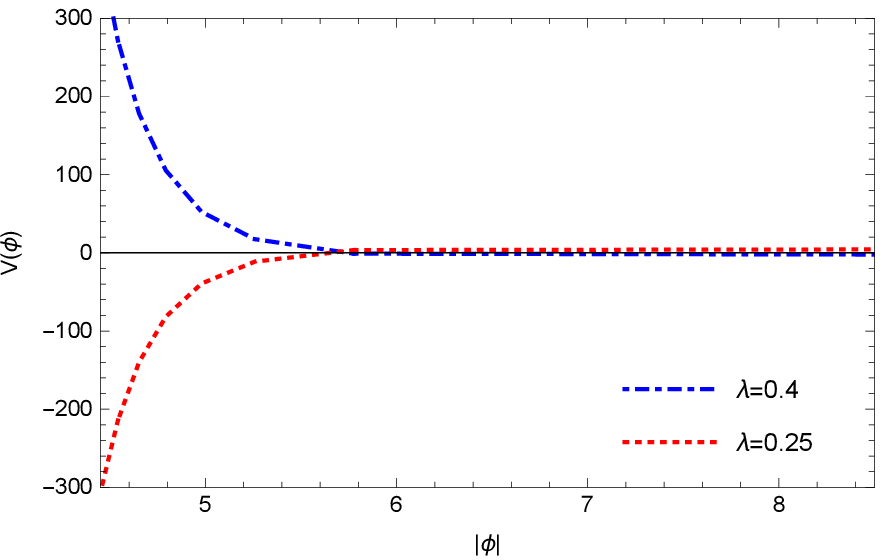}\\
	\end{minipage}
\caption{Scalar field and Potential function is plotted for flat case ($k=0$)}
\label{fig3}
\end{figure}
with graphical representation in fig. \ref{fig3}. One may note that `$\cosh$' type potential is appropriate for inflationary era according to Planck data \cite{Akrami:2018odb}.


\section{Thermodynamical Analysis}
For the second part of our investigation we shall now analyze the thermodynamical behaviour of this non-singular era of evolution. In particular, we shall examine the validity of the generalized second law of thermodynamics (GSLT) and thermodynamical equilibrium (TE) for the emergent era with Universe bounded by apparent horizon (event horizon does not exist in emergent era).

The FLRW metric can have the following (2+2)-decomposition
\begin{equation}
ds^2=h_{ij}(x^i)dx^idx^j+R^2 d\Omega^2_2
\end{equation}
where\vspace{-.5cm} $$h_{ij}=diag\left(-1,\frac{a^2}{1-kr^2}\right);~i,j=0,1$$

is the metric on the $2$-space, normal the 2-sphere and $R=ar$ is the area radius, a scalar field in the normal 2D space. Another scalar 
\begin{equation}
\chi(x)=h^{ij}(x)\partial_i R \partial_j R=1-\left(H^2+\frac{k}{a^2}\right) R^2
\end{equation}
on the normal $(r,t)$-plane defined the surface gravity (on the horizon) as
\begin{equation}
K_H=-\frac{1}{2}\frac{\partial \chi}{\partial R}\bigg|_{R=R_H}=R_H\left(H^2+\frac{k}{a^2}\right)
\end{equation}
and $\chi=0$ identifies a null surface known as apparent horizon having area radius
\begin{equation}
R_A=\frac{1}{\sqrt{H^2+\frac{k}{a^2}}}
\end{equation}

Hence the usual Hawking temperature on the horizon gives \cite{Chakraborty:2012cw}
\begin{equation}
T_H=\frac{|K_H|}{2\pi}=\frac{R_H}{2\pi R_A^2}
\end{equation}

If $S_H$ and $S_f$ denote the entropy of the horizon and that of the cosmic fluid bounded by the horizon then GSLT and TE demand \cite{Chakraborty:2014oya} 
\begin{equation}
(i) \dot{S_H}+\dot{S_f}>0\mbox{~ and~ } (ii) \ddot{S_H}+\ddot{S_f}<0
\end{equation}

Now, the entropy on an horizon (having area radius $R_H$) can be determined from the first law of thermodynamics (i.e. Clausius relation)
\begin{equation}
T_H dS_H=\delta Q=-dE=4\pi R_H^2 (\rho+p) H dt
\end{equation}
as\vspace{-.5cm}
\begin{equation}
\dot{S_H}=\frac{4\pi R_H^2 H (\rho+p)}{T_H}
\end{equation}

On the other hand, using Gibbs relation
\begin{equation}
T_f dS_f=dE+pdV
\end{equation}
the explicit form of the variation of the entropy of the cosmic fluid has the expression
\begin{equation}
\dot{S_f}=\frac{4\pi R_H^2(p+\rho)}{T_f}\left(\dot{R_H}-HR_H\right)
\end{equation}
where $V=\frac{4}{3}\pi R_H^3$ is the volume of the cosmic fluid, $E=\rho V$ is the total energy and $T_f$ is the temperature of the fluid. Hence the total entropy variation is given by
\begin{equation}
\dot{S_T}=\dot{S_H}+\dot{S_f}=\frac{4\pi R_H^2(p+\rho)}{T_f}\left(\dot{R_H}-H(R_H-1)\right)
\end{equation}
where $T_f=T_H$ is assumed for equilibrium configuration.

 Using the 2nd Friedmann equation 
\begin{equation}
\left(\dot{H}-\frac{k}{a^2}\right)=-4\pi(p+\rho)
\end{equation}
and equation (27) for the Horizon temperature, the total entropy variation takes the form;
\begin{equation}
\dot{S_T}=-2R_A^2R_H\left(\dot{H}-\frac{k}{a^2}\right)\left[\dot{R_H}-H(R_H-1)\right]
\end{equation}

Differentiating once more one gets
\begin{eqnarray}
\ddot{S_T}&=&-2\bigg[\left(2R_A\dot{R_A}R_H+R_A^2\dot{R_H}\right)\left(\dot{H}-\frac{k}{a^2}\right)\bigg\{\dot{R_H}-H\left(R_H-1\right)\bigg\}\nonumber\\&~&
~~~~~+R_A^2R_H\left(\ddot{H}+\frac{2k}{a^2}H\right)\bigg\{\dot{R_H}-H\left(R_H-1\right)\bigg\}\nonumber\\&~&
~~~~~+R_A^2R_H\left(\dot{H}-\frac{k}{a^2}\right)\bigg\{\ddot{R_H}-\dot{H}\left(R_H-1\right)-\dot{R_H}H\bigg\}\bigg]
\end{eqnarray}

As at the emergent era $R_A$ is constant, $H=0$, $\dot{H}=0$ so the expressions for total entropy variation take the form (horizon is distinct from apparent horizon)
\begin{eqnarray}
\dot{S_T}&=&\frac{R_A^2k}{a_0^2}\frac{d}{dt}\left(R_H^2\right)\nonumber\\
\mbox{and~~}\ddot{S_T}&=&\frac{R_A^2k}{a_0^2}\frac{d^2}{dt^2}\left(R_H^2\right)
\end{eqnarray} 

So the validity of GSLT and thermodynamical equilibrium can be constrained in the following table.
\begin{table}[h]
	\caption{\label{t2}Condition for GSLT and thermal equilibrium}
	\begin{tabular}{|c|c|c|}
		\hline
	Curvature scalar&GSLT($\dot{S_T}>0$)&Thermodynamical equilibrium($\ddot{S_T}<0$)\\\hline
	$k=0$&not satisfied&no\\\hline
		$k=+1$&$\frac{d}{dt}\left(R_H^2\right)>0$&$\dfrac{d^2}{dt^2}\left(R_H^2\right)<0$\\\hline
		$k=-1$&$\frac{d}{dt}\left(R_H^2\right)<0$&$\frac{d^2}{dt^2}\left(R_H^2\right)>0$\\\hline
	\end{tabular}
\end{table}

The table \ref{t2} shows that both GSLT and thermodynamical equilibrium is satisfied for closed model (i.e. $k=+1$) if the area radius increases in a decelerated way while for open model (i.e. $k=-1$) if the area radius decreases in an accelerated way.

In the emergent era the area radius at the apparent horizon is given by 
\begin{eqnarray}
R_A&=& \frac{a_0}{\sqrt{|k|}}~,~ k\neq0\nonumber\\&=&0~~~~~~,~k=0\nonumber
\end{eqnarray}

Thus both $\dot{S_T}$ and $\ddot{S_T}$ vanish for Universe bounded by apparent horizon in the emergent era. Hence GSLT is not obeyed and one can not predict about thermodynamical equilibrium in non-singular emergent phase for Universe bounded by the apparent horizon.
\section{Summary \& Conclusion}
A detailed study of the emergent phase of evolution is considered in the present work for HL gravity theory. In the background of homogeneous and isotropic space-time model the occurrence of  emergent era restricts the perfect fluid as exotic type both for flat and closed model while for open model normal perfect fluid is sufficient for non-singular epoch. In equivalent scalar field description, the emergent scenario demands ghost type scalar field for flat and closed model; and normal scalar field for open model of the Universe. However it is to be noted that this scalar field description of the barotropic fluid model is not unique in general \cite{Gao:2009me} and even it may encounter few problems \cite{Linder:2008ya}. Further, Baryogenesis occurs separately from any kind of emergent model \cite{Cline:2006ts} and it is independent of any extended/modified theory of gravity \cite{Capozziello:2019cav}. Finally, from thermodynamical analysis it is found that GSLT is not valid and there is no thermodynamical equilibrium for this static phase at the apparent horizon while no definite conclusion is possible at any other horizon.

For future work, it will be interesting to consider singularity free cosmic solution which may evolve continuously to inflationary era.
\section*{Acknowledgment}
 The author A.B. acknowledges UGC for awarding JRF (ID.1207/CSIRNETJUNE2019) and S.C. thanks Science and Engineering Research Board (SERB), India for awarding MATRICS Research Grant support (FileNo.MTR/2017/000407) and RUSA 2.0 of Jadavpur Univerity.


\begin{thebibliography}{50}
	\bibitem{Harrison:1967zz} 
	E.~R.~Harrison,
	Mon.\ Not.\ Roy.\ Astron.\ Soc.\  {\bf 137}, 69 (1967).
	\bibitem{Ellis:2002we} 
	G.~F.~R.~Ellis and R.~Maartens,
	Class.\ Quant.\ Grav.\  {\bf 21}, 223 (2004)
	\bibitem{Ellis:2003qz} 
	G.~F.~R.~Ellis, J.~Murugan and C.~G.~Tsagas,
	Class.\ Quant.\ Grav.\  {\bf 21}, no. 1, 233 (2004)
	\bibitem{Mukherjee:2005zt} 
	S.~Mukherjee, B.~C.~Paul, S.~D.~Maharaj and A.~Beesham,
	gr-qc/0505103.
	\bibitem{Mukherjee:2006ds} 
	S.~Mukherjee, B.~C.~Paul, N.~K.~Dadhich, S.~D.~Maharaj and A.~Beesham,
	Class.\ Quant.\ Grav.\  {\bf 23}, 6927 (2006)
	\bibitem{Mulryne:2005ef} 
	D.~J.~Mulryne, R.~Tavakol, J.~E.~Lidsey and G.~F.~R.~Ellis,
	Phys.\ Rev.\ D {\bf 71}, 123512 (2005)
	\bibitem{Banerjee:2007qi} 
	A.~Banerjee, T.~Bandyopadhyay and S.~Chakraborty,
	Grav.\ Cosmol.\  {\bf 13}, 290 (2007)
	Gen.\ Rel.\ Grav.\  {\bf 40}, 1603 (2008)
	\bibitem{Nunes:2005ra} 
	N.~J.~Nunes,
	Phys.\ Rev.\ D {\bf 72}, 103510 (2005)
	\bibitem{Lidsey:2006md} 
	J.~E.~Lidsey and D.~J.~Mulryne,
	Phys.\ Rev.\ D {\bf 73}, 083508 (2006)
	\bibitem{Debnath:2008nu} 
	U.~Debnath,
	Class.\ Quant.\ Grav.\  {\bf 25}, 205019 (2008)
	\bibitem{Paul}
	B.~C.~Paul and S.~Ghose,
	Gen.\ Relt.\ Grav.{\bf 42}, 795 (2010)
	\bibitem{Debnath:2011qi} 
	U.~Debnath and S.~Chakraborty,
	Int.\ J.\ Theor.\ Phys.\  {\bf 50}, 2892 (2011)
	\bibitem{Mukerji:2011wq} 
	S.~Mukerji, N.~Mazumder, R.~Biswas and S.~Chakraborty,
	Int.\ J.\ Theor.\ Phys.\  {\bf 50}, 2708 (2011)
	\bibitem{Labrana:2011np} 
	P.~Labrana,
	Phys.\ Rev.\ D {\bf 86}, 083524 (2012)
\bibitem{Chakraborty:2014ora} 
S.~Chakraborty,
Phys.\ Lett.\ B {\bf 732}, 81 (2014).
\bibitem{delCampo:2011aa} 
S.~del Campo, R.~Herrera and D.~Pavon,
Phys.\ Lett.\ B {\bf 707}, 8 (2012).
\bibitem{Kiritsis:2009sh} 
E.~Kiritsis and G.~Kofinas,
Nucl.\ Phys.\ B {\bf 821}, 467 (2009).
\bibitem{Fonseca-Neto:2013rna} 
J.~B.~Fonseca-Neto, A.~Y.~Petrov and M.~J.~Reboucas,
Phys.\ Lett.\ B {\bf 725}, 412 (2013).
\bibitem{Jing:2010cy} 
M.~Wang, S.~Chen and J.~Jing,
Phys.\ Lett.\ B {\bf 695}, 401 (2011).
\bibitem{Myung:2010dv} 
Y.~S.~Myung,
Phys.\ Lett.\ B {\bf 690}, 534 
;\ {\bf 685}, 318 
;\ {\bf 684}, 158 (2010)
;\ {\bf 679}, 491
;\ {\bf 678}, 127 (2009).
\bibitem{Leon:2019mbo}
G.~Leon and A.~Paliathanasis,
Eur. Phys. J. C \textbf{79} (2019) no.9, 746.
\bibitem{Heydarzade:2015hra} 
Y.~Heydarzade, M.~Khodadi and F.~Darabi,
Theor.\ Math.\ Phys.\  {\bf 190}, no. 1, 130 (2017).
\bibitem{Pourhassan:2017qhq} 
B.~Pourhassan, S.~Upadhyay, H.~Saadat and H.~Farahani,
Nucl.\ Phys.\ B {\bf 928}, 415 (2018).
\bibitem{Cai:2009qs} 
R.~G.~Cai, L.~M.~Cao and N.~Ohta,
Phys.\ Lett.\ B {\bf 679}, 504 (2009).
\bibitem{Horava:2009uw} 
P.~Horava,
Phys.\ Rev.\ D {\bf 79}, 084008 (2009).
\bibitem{Sotiriou:2010wn}
T.~P.~Sotiriou,
J. Phys. Conf. Ser. \textbf{283} (2011), 012034.
\bibitem{Luongo:2018oil}
O.~Luongo, M.~Muccino and H.~Quevedo,
Phys. Dark Univ. \textbf{25} (2019), 100313.
\bibitem{Akrami:2018odb}
Y.~Akrami \textit{et al.} [Planck],
Astron. Astrophys. \textbf{641} (2020), A10.
\bibitem{Chakraborty:2012cw} 
S.~Chakraborty,
Phys.\ Lett.\ B {\bf 718}, 276 (2012).
\bibitem{Chakraborty:2014oya} 
S.~Chakraborty and S.~Saha,
Phys.\ Rev.\ D {\bf 90}, no. 12, 123505 (2014).
\bibitem{Gao:2009me} 
C.~Gao, M.~Kunz, A.~R.~Liddle and D.~Parkinson,
Phys.\ Rev.\ D {\bf 81}, 043520 (2010)
\bibitem{Linder:2008ya}
E.~V.~Linder and R.~J.~Scherrer,
Phys. Rev. D \textbf{80} (2009), 023008
\bibitem{Cline:2006ts}
J.~M.~Cline,
[arXiv:hep-ph/0609145 [hep-ph]].
\bibitem{Capozziello:2019cav}
S.~Capozziello, R.~D'Agostino and O.~Luongo,
Int. J. Mod. Phys. D \textbf{28} (2019) no.10, 1930016
\end{thebibliography}
\end{document}